# High pressure, high temperature molecular doping of nanodiamond

**Authors:** M. Crane[1], A. Petrone[2], R. Beck[2], M. Lim[3], X. Zhou[3], X. Li[2], R. Stroud[4], P. Pauzauskie[1,3,5]*

**Affiliations:**

[1]Department of Chemical Engineering, University of Washington, Seattle, WA 98195-1750, United States

[2]Department of Chemistry, University of Washington, Seattle, WA, 98195-1700, United States

[3]Department of Materials Science and Engineering, University of Washington, Seattle, WA 98195-2120, United States

[4]Materials Science and Technology Division, Naval Research Laboratory, Washington, DC 20375, United States

[5]Physical and Computational Sciences Directorate, Pacific Northwest National Laboratory, Richland, WA 99352, United States

*peterpz@uw.edu

**Abstract**: The development of color centers in diamond as the basis for emerging quantum technologies has been limited by the need for ion implantation to create the appropriate defects. We present a versatile method to dope diamond without ion implantation, by synthesis of a doped amorphous carbon precursor and transformation at high temperatures and high pressures. To explore this bottom-up method for color center generation, we rationally create silicon-vacancy defects in nanodiamond and investigate them for optical pressure metrology. In addition, we show that this process can generate noble gas defects within diamond from the typically-inactive argon pressure medium, which may explain the hysteresis effects observed in other high pressure experiments and the presence of noble gases in some meteoritic nanodiamonds. Our results illustrate a general method to produce color centers in diamond, and may enable the controlled generation of designer defects.

**Main Text:** The characterization and manipulation of dopants in diamond has generated a wide range of applications spanning quantum computing, sensing, and cryptography (*1–3*), the determination of interstellar origin in meteoritic samples (*4–6*), characterization of the earth's mantle(*7, 8*), and biolabeling (*9*), due to the remarkable properties of the diamond host. The dense diamond lattice exhibits a negligible immune response, maintains a wide bandgap, and, notably, restricts heteroatom defect diffusion at temperatures far above the diamond-graphite phase line at atmospheric pressure. For example, a common defect in diamond, substitutional nitrogen, does not diffuse at temperatures below 2000°C (*10*). In quantum sensing applications, this low diffusion coefficient enables the reliable use single-defects like the nitrogen vacancy center ($NV^-$) to optically measure local spatiotemporal variations, which modify the defect's spin precession rate, without fear of color center migration over long time scales (*1*). Similar applications in quantum cryptography have been proposed for the negatively-charged, silicon-divacancy ($SiV^-$) center. Because diffusion doping is not practical in diamond at ambient pressure, ion implantation is typically used to incorporate heteroatomic defects. This process relies on Poisson statistics, SRIM calculations, and masking techniques to control color center generation in chemical vapor deposition (CVD) diamond substrates (*2*). However, ion implantation also creates significant lattice damage, induces fragmentation of ions, and cannot deterministically produce polyatomic defects. Due to these challenges, progress in single-defect applications often occurs by bulk defect production with implantation followed by confocal-scanning searches for ideal color centers.

An alternative method for doped, diamond synthesis is high pressure, high temperature (HPHT) equilibrium phase conversion (*11*, *12*). While HPHT processes have produced doped diamonds, the rational formation of heteroatomic defects has remained elusive (*13*). In addition, HPHT experiments, including diamond syntheses, conventionally employ noble gas pressure media, which, if incorporated into the lattice, have been proposed as defects for quantum sensing. However, to date, noble gas defect formation, such as xenon-related dopants, has been restricted to ion implantation (*14*). Despite its nearly-ubiquitous role in high pressure experiments, noble gas pressure media is widely considered to be inert, and there are no studies regarding the conditions that lead to incorporation within the diamond lattice at HPHT conditions (*15*, *16*).

To overcome diamond's low diffusion coefficient and to study the incorporation of noble gas dopants without ion implantation, we propose a bottom-up methodology to dope diamond by first synthesizing a doped amorphous carbon precursor and then converting it to diamond at HPHT conditions in a noble gas environment. This allows us to simultaneously integrate the desired dopant into carbon while it is thermodynamically stable with traditional synthetic chemistry techniques, rather than rely on ion implantation into a metastable diamond substrate, and investigate noble gas incorporation at HPHT.

Here, we probe these hypotheses by synthesizing a nanostructured carbon aerogel precursor with a controlled chemical composition, and subjecting it to HPHT conditions in a laser-heated diamond anvil cell (DAC) with an argon pressure medium, as shown in Fig. 1 (*13*, *17*). Bright-field transmission electron microscopy (BF-TEM) and selected area electron diffraction (SAED) in Fig. 1 demonstrate that the aerogel consists of 6.8 ± 1.9 nm radius amorphous carbon grains. We tuned the chemical composition of the aerogel grains by adding tetraethylorthosilicate (TEOS) molecules directly to the mixture as it gelled (Fig. 1a). Energy dispersive X-ray spectroscopy (EDS) confirmed that silicon dopants were incorporated throughout the carbon precursor material. To synthesize diamond, we placed the doped carbon precursor into a diamond anvil cell and condensed solid argon within the high-pressure chamber to infiltrate the microstructure of the aerogel (Fig. 1b). We subsequently pressurized the cell above 20 GPa to thermodynamically favor diamond formation and drove grain growth by heating above 2000 K with a near-infrared, continuous laser (Fig. 1c) (*11*).

To characterize the recovered material, we examined BF-TEM, SAED, and electron energy loss spectroscopy (EELS). We found that the recovered material exhibited nanocrystalline domains with interlayer distances corresponding to cubic diamond (Fig. 2b, Fig. S1). The nanodiamond grain sizes ranged from 1 - 200 nm (Fig. S2), indicating that significant carbon diffusion occurs during HPHT synthesis, which was likely enhanced by the high synthesis temperatures that surpass the melting point of argon at 20 GPa (1580 K) (*18*). The carbon K-edge EELS spectra of pure diamond has a characteristic near-edge structure with prominent $\sigma^*$ peak at 290 eV a dip at 302.5 eV (*19*). The carbon K-edge spectrum of the recovered material contained both features, further indicating that the HPHT treatment formed cubic diamond, as well as a small pre-edge peak at 285 eV. This pre-edge feature corresponds to a $\pi^*$ excitation associated with $sp^2$ carbon (*5, 19*). As observed in previous HPHT and CVD experiments, this $sp^2$ carbon likely stems from nanodiamond surface reconstruction and incomplete sample heating due to the self-limiting absorption of amorphous carbon as it converts to diamond (*13*). Low energy (Fig. S4) loss data and Raman scattering (Fig. S3) from the recovered material similarly indicate the prevalence of $sp^3$ carbon in a diamond structure with a small amount of $sp^2$ carbon (*20*).

EDS and EELS allow us to measure the chemical composition of the recovered nanodiamond and confirm the presence of dopants, including nitrogen, silicon, and argon. The Z-contrast of high-angle annular dark field scanning transmission electron microscopy (STEM-HAADF) images identify individual atoms and clusters (Fig. 2a). Combined, these data unambiguously demonstrate that silicon dopants added to the carbon aerogel precursor remain in and/or on the nanodiamond product after heating, despite significant grain growth. The presence of argon within the recovered material, despite decompression to atmospheric pressure, transfer to a TEM grid, and analysis at ultrahigh vacuum conditions, demonstrates robust incorporation within the diamond lattice, rather than surface adsorption.

We observed argon in all recovered samples synthesized at a range of pressures and temperatures from 20 to 25 GPa and 1800 to 3000 K. While other reports have demonstrated the effect of noble gas pressure media on samples at elevated pressures, such as partitioning of helium in $SiO_2$ (*21*), this is the first confirmation of noble gas incorporation and recovery from HPHT (*22*). These results suggest that the aerogel structure allows argon to incorporate within its micropores during compression and that grain growth during laser heating traps these atoms within the lattice. For optoelectronic color center applications in diamond, this represents a new methodology for the incorporation of noble gas defects, e.g. xenon, for quantum computing and sensing (*23*). In addition, noble gas pressure media are almost exclusively used in HPHT experiments because they remain hydrostatic to high pressures and are chemically and physically inert (*24*). The incorporation of noble gas pressure media into materials at HPHT conditions challenges the view of complete inactivity, and could explain hysteresis effects in prior DAC experiments (*25*). It could also provide an explanation for how noble gas atoms are incorporated into nanodiamonds in astrophysical environments (*5*).

The PL spectra of all the recovered material contain optically-active color centers from NV centers (Fig. 3). The shoulders at 575 and 637 nm and the broad feature centered at 700 nm are uniquely characteristic of $NV^0$ and $NV^-$ zero phonon lines (ZPLs) and phonon side bands, which have been observed in multiple HPHT reports due to atmospheric $N_2$ incorporation (*13*, *26*). PL signal from the argon is neither observed nor expected (*27*). However, the silicon-doped carbon aerogel contains a peak at 739 nm, corresponding to the $SiV^-$ color center, that is not present in undoped carbon aerogel (*28*). These observations confirm that dopants added to the carbon aerogel precursor persist within the nanodiamond lattice and provide a new mechanism for engineering dopants in diamond. Both SiV and NV are lower energy states than their substitutional silicon and nitrogen counterparts due to lattice relaxation, as observed in Jahn-Teller distortion of the lattice (*29*, *30*). Unlike ion implantation, which requires annealing to drive vacancy diffusion to activate incorporated heteroatoms, optically-active color centers form immediately upon HPHT conversion to diamond. This suggests that as the diamond lattice forms around the heteroatomic silicon and nitrogen atoms, the lowest energy color center structure forms immediately. While silicon atoms have been doped into diamond before, the process involved ion implantation or complete chemical breakdown in a plasma which limited controllable heteroatom defect formation (*2*). This bottom-up approach illustrates the possibility of a new doping paradigm for diamond where molecular dopants can be designed with the precise heteroatomic stoichiometry and three-dimensional stereochemistry to create a wide range of multifunctional polyatomic point defects.

As discussed above, color centers in diamond are attractive materials for optical sensing applications due to their high stability in the chemically inert diamond lattice (*26*). Due to its narrow linewidth, the SiV$^-$ center may act as a high-resolution pressure sensor. To evaluate the defect for optical pressure metrology and illustrate the efficacy of bottom-up doping, we collected pressure-dependent photoluminescence (Fig. 4a) spectra, which reveals a 0.98 meV/GPa slope from 0 to 25 GPa. We employed ab initio calculations to model the pressure-dependence of the SiV$^-$ by fully simulating a nearly-spherical $C_{119}SiH_{104}$ nanodiamond (~1.2 nm in diameter) containing a SiV$^-$ defect under the effect of the uniform hydrostatic pressure with density functional theory using the Gaussian electronic structure package (*31*). These theoretical results predict a 0.8 meV/GPa shift from 0 to 25 GPa, in close agreement with experimental observations. Extending the simulation up to 140 GPa (Fig. S5) demonstrates the viability of optical pressure metrology with SiV$^-$ to high pressures. To date, the high quantum efficiency, narrow linewidth d-d transitions of $Cr^{3+}$ in alumina (ruby) have made it the nearly-ubiquitous choice to measure pressure at in high pressure DAC experiments (*32*). However, ruby undergoes a phase transition at 94 GPa at 1300 °C (*33*), making it unsuitable for the next generation of HPHT experiments, which have recently reached the terapascal range (*34*). On the other hand, diamond is the thermodynamically-stable polytype of carbon at all temperatures pressures above 1 GPa until melting (*11*). The lack of phase transformation implies that SiV$^-$-doped nanodiamond may succeed at conditions where ruby fails.

The rational incorporation of silicon by chemically doping the carbon precursor with TEOS and argon by employing an argon pressure media into nanodiamond illustrate the potential impact of this doping methodology for doped nanodiamond applications, like pressure metrology, without ion implantation. Rather than sequentially synthesizing diamond, implanting substitutional heteroatoms, annealing vacancy center, and conducting confocal searches for color centers, the HPHT conversion of doped carbon can directly form color centers. For single-defect applications, this research opens the door to the incorporation of more complex defects into diamond with structures defined by the chemical dopant added into the carbon precursor. If diamond nucleates prior to dissociation of the dopant, defects could be added with chemical precision limited only by molecular synthesis. For extraterrestrial nanodiamonds, where dopants in diamond are used to fingerprint the presolar and interstellar environment, this demonstration unveils the DAC as a tool to study HPHT doping that could occur in astrophysical environments (*4, 5*). Given the prevalence of noble gas pressure media, these results have broad implications for high pressure experiments, where, to date, noble gasses have been considered inert (*21, 24*).

**References and Notes**

1. F. Dolde *et al.*, Nanoscale Detection of a Single Fundamental Charge in Ambient Conditions Using the NV$^-$ Center in Diamond. *Phys. Rev. Lett.* **112**, 097603 (2014).

2. S. Pezzagna, D. Rogalla, D. Wildanger, J. Meijer, A. Zaitsev, Creation and nature of optical centres in diamond for single-photon emission—overview and critical remarks. *New J. Phys.* **13**, 035024 (2011).

3. N. Aslam *et al.*, Nanoscale nuclear magnetic resonance with chemical resolution. *Science*. **357**, 67–71 (2017).

4. A. B. Verchovsky *et al.*, C, N, and Noble Gas Isotopes in Grain Size Separates of Presolar Diamonds from Efremovka. *Science*. **281**, 1165–1168 (1998).

5.  R. M. Stroud, M. F. Chisholm, P. R. Heck, C. M. O. Alexander, L. R. Nittler, Supernova Shock-wave-induced CO-formation of Glassy Carbon and Nanodiamond. *Astrophys. J. Lett.* **738**, L27 (2011).

6.  R. S. Lewis, T. Ming, J. F. Wacker, E. Anders, E. Steel, Interstellar diamonds in meteorites. *Nature*. **326**, 160 (1987).

7.  O. Tschauner *et al.*, Ice-VII inclusions in diamonds: Evidence for aqueous fluid in Earth's deep mantle. *Science*. **359**, 1136–1139 (2018).

8.  F. Nestola *et al.*, CaSiO3 perovskite in diamond indicates the recycling of oceanic crust into the lower mantle. *Nature*. **555**, 237–241 (2018).

9.  V. N. Mochalin, O. Shenderova, D. Ho, Y. Gogotsi, The properties and applications of nanodiamonds. *Nat. Nanotechnol.* **7**, 11–23 (2012).

10. R. M. Chrenko, R. E. Tuft, H. M. Strong, Transformation of the state of nitrogen in diamond. *Nature*. **270**, 141–144 (1977).

11. F. P. Bundy *et al.*, The pressure-temperature phase and transformation diagram for carbon; updated through 1994. *Carbon*. **34**, 141–153 (1996).

12. S. Choi *et al.*, Varying temperature and silicon content in nanodiamond growth: effects on silicon-vacancy centres. *Sci. Rep.* **8**, 3792 (2018).

13. P. J. Pauzauskie *et al.*, Synthesis and characterization of a nanocrystalline diamond aerogel. *Proc. Natl. Acad. Sci.* **108**, 8550–8553 (2011).

14. R. Sandstrom *et al.*, Optical properties of implanted Xe color centers in diamond. *Opt. Commun.* **411**, 182–186 (2018).

15. S. Basu, A. P. Jones, A. B. Verchovsky, S. P. Kelley, F. M. Stuart, An overview of noble gas (He, Ne, Ar, Xe) contents and isotope signals in terrestrial diamond. *Earth-Sci. Rev.* **126**, 235–249 (2013).

16. A. P. Koscheev, M. D. Gromov, R. K. Mohapatra, U. Ott, History of trace gases in presolar diamonds inferred from ion-implantation experiments. *Nature*. **412**, 615–617 (2001).

17. R. W. Pekala, Organic aerogels from the polycondensation of resorcinol with formaldehyde. *J. Mater. Sci.* **24**, 3221–3227 (1989).

18. J. H. Carpenter, S. Root, K. R. Cochrane, D. G. Flicker, T. R. Mattsson, "Equation of state of argon: experiments on Z, density functional theory (DFT) simulations, and wide-range model" (Unlimited Release SAND2012-7991, Sandia National Laboratories, Sandia National Laboratories, 2012), pp. 1–56.

19. R. F. Egerton, *Electron Energy-Loss Spectroscopy in the Electron Microscope* (Plenum Press, New York, NY, ed. 2, 1996).

20. A. C. Ferrari, J. Robertson, Raman spectroscopy of amorphous, nanostructured, diamond–like carbon, and nanodiamond. *Philos. Trans. R. Soc. Lond. Math. Phys. Eng. Sci.* **362**, 2477–2512 (2004).

21. G. Shen *et al.*, Effect of helium on structure and compression behavior of SiO2 glass. *Proc. Natl. Acad. Sci.* **108**, 6004–6007 (2011).

22. K. Fukunaga, J. Matsuda, K. Nagao, M. Miyamoto, K. Ito, Noble-gas enrichment in vapour-growth diamonds and the origin of diamonds in ureilites. *Nature*. **328**, 141–143 (1987).

23. V. A. Martinovich, A. V. Turukhin, A. M. Zaitsev, A. A. Gorokhovsky, Photoluminescence spectra of xenon implanted natural diamonds. *J. Lumin.* **102–103**, 785–790 (2003).

24. W. A. Caldwell *et al.*, Structure, Bonding, and Geochemistry of Xenon at High Pressures. *Science*. **277**, 930–933 (1997).

25. R. Boehler, N. von Bargen, A. Chopelas, Melting, thermal expansion, and phase transitions of iron at high pressures. *J. Geophys. Res. Solid Earth*. **95**, 21731–21736 (1990).

26. M. W. Doherty *et al.*, Electronic Properties and Metrology Applications of the Diamond $NV^-$ Center under Pressure. *Phys. Rev. Lett.* **112**, 047601 (2014).

27. A. M. Zaitsev, *Optical Properties of Diamond* (Springer Berlin Heidelberg, Berlin, Heidelberg, 2001; http://link.springer.com/10.1007/978-3-662-04548-0).

28. T. Feng, B. D. Schwartz, Characteristics and origin of the 1.681 eV luminescence center in chemical-vapor-deposited diamond films. *J. Appl. Phys.* **73**, 1415–1425 (1993).

29. C. Hepp *et al.*, Electronic Structure of the Silicon Vacancy Color Center in Diamond. *Phys. Rev. Lett.* **112**, 036405 (2014).

30. A. Lenef, S. C. Rand, Electronic structure of the N-V center in diamond: Theory. *Phys. Rev. B*. **53**, 13441–13455 (1996).

31. M. J. Frisch *et al.*, *Gaussian 16* (Wallingford, CT, 2016).

32. H. K. Mao, J. Xu, P. M. Bell, Calibration of the ruby pressure gauge to 800 kbar under quasi-hydrostatic conditions. *J. Geophys. Res. Solid Earth*. **91**, 4673–4676 (1986).

33. J.-F. Lin *et al.*, Crystal structure of a high-pressure/high-temperature phase of alumina by in situ X-ray diffraction. *Nat. Mater.* **3**, 389–393 (2004).

34. N. Dubrovinskaia *et al.*, Terapascal static pressure generation with ultrahigh yield strength nanodiamond. *Sci. Adv.* **2**, e1600341 (2016).

**Acknowledgments:** We sincerely thank E. Abramson, B. Smith, and P. Meisenheimer for their advice and assistance with high pressure experiments. This research was made possible by a CAREER Award from the National Science Foundation (Award #1555007), starting funding from the University of Washington, and the Materials Research Science and Engineering Center (MDR-171997). M. Crane gratefully acknowledges support from the Department of Defense through a National Defense Science and Engineering Graduate Fellowship (NDSEG) program, the Microanalysis Society of America through a Joseph Goldstein Scholar Award, and R. M. Stroud for mentoring at the NRL. P. Pauzauskie gratefully acknowledges support from both the US Department of Energy's Pacific Northwest National Laboratory (PNNL) and the Materials and Simulation Across Scales (MS3) Initiative, a Laboratory Directed Research and Development (LDRD) program at the PNNL. X. Li gratefully acknowledges support from National Science Foundation (CHE-1565520 and CHE-1464497). This work was facilitated by advanced computational, storage, and networking infrastructure provided by the Hyak supercomputer system at University of Washington, funded by the Student Technology Fee.

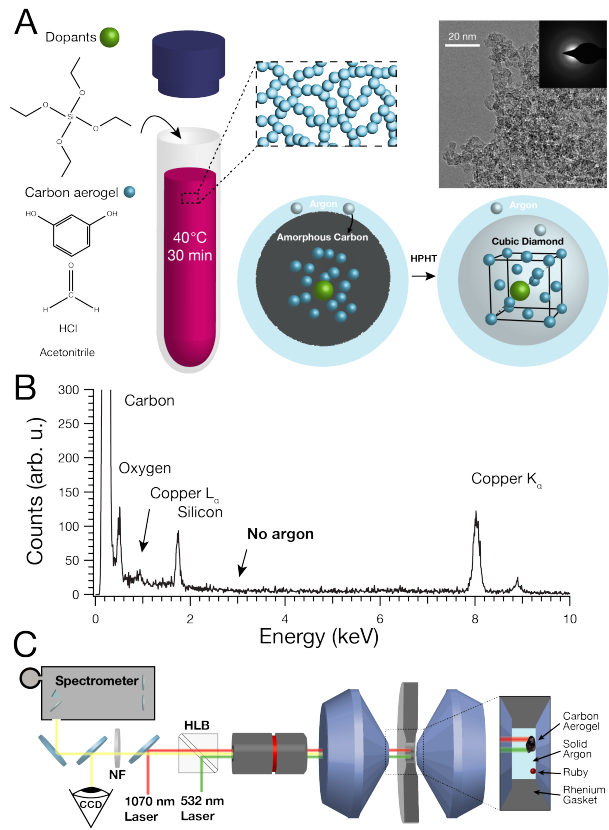

**Fig. 1. Carbon precursor doping mechanism and characterization.** (**A**) Schematic representing the synthesis and doping of carbon aerogels, including BF-TEM image with SAED inset. Dopants are introduced alongside resorcinol and formaldehyde incorporate within the carbon aerogel grains. Upon conversion to diamond at high pressure and high temperature, dopants remain inside the diamond lattice as color centers. (**B**) EDS spectra of the carbon aerogel as synthesized show only the presence of carbon, silicon, and oxygen. Copper signal comes from the TEM grid. (**C**) Schematic showing a 1070 nm heating laser or polarized 532 nm Raman and photoluminescence laser focused into the pressurized diamond anvil cell, which is loaded with a carbon aerogel precursor, ruby for pressure measurements, and a solid argon pressure media, contained by a rhenium gasket.

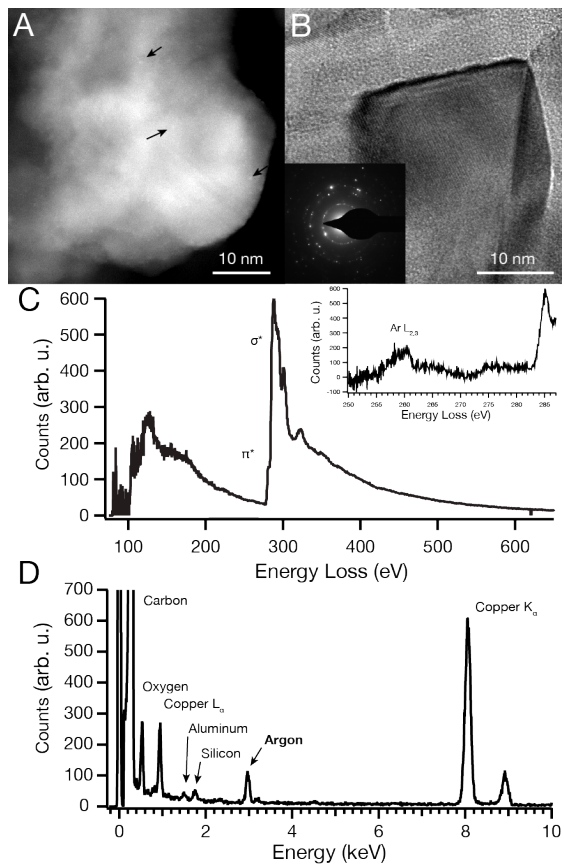

**Fig. 2. Structural and chemical characterization after HPHT synthesis.** (**A** and **B**) STEM-HAADF and BF-TEM illustrate the microstructure of the recovered diamond material. Arrows in the STEM-HAADF image point out example impurity atoms (mostly Ar and Si), and the inset in panel (B) contains SAED, corresponding to diamond. In addition, (B) exhibits 2.08 Å lattice spacings. Fig. S1 details the d-spacing assignments. (**C** and **D**) STEM-EELS and STEM-EDS of the region displayed in (A) showing the silicon L-edge and carbon K-edge, and elemental composition, respectively. The small concentration of aluminum likely comes from trace

amounts of ruby during laser heating. The Cu peak is from the sample grid and STEM pole piece. The inset in panel (C) displays the carbon pre-edge features and the argon $L_{2,3}$ peak.

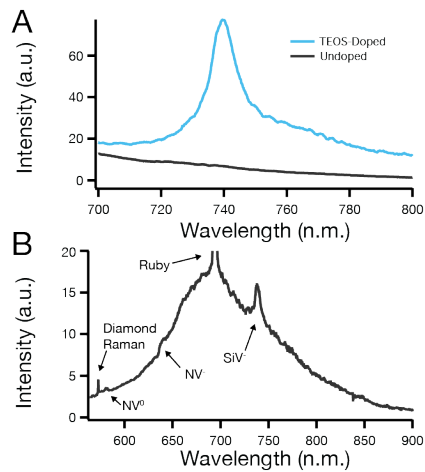

**Fig. 3. Photoluminescence of color center.** (**A**) High resolution photoluminescence spectra of the SiV⁻ region comparing TEOS-doped and undoped carbon aerogels. (**B**) Photoluminescence and Raman scattering of recovered nanodiamond synthesized from the TEOS-doped carbon aerogel after depressurization and removal from the DAC. Labels denote diamond Raman scattering, and $NV^0$, $NV^-$, and SiV⁻ color center ZPLs.

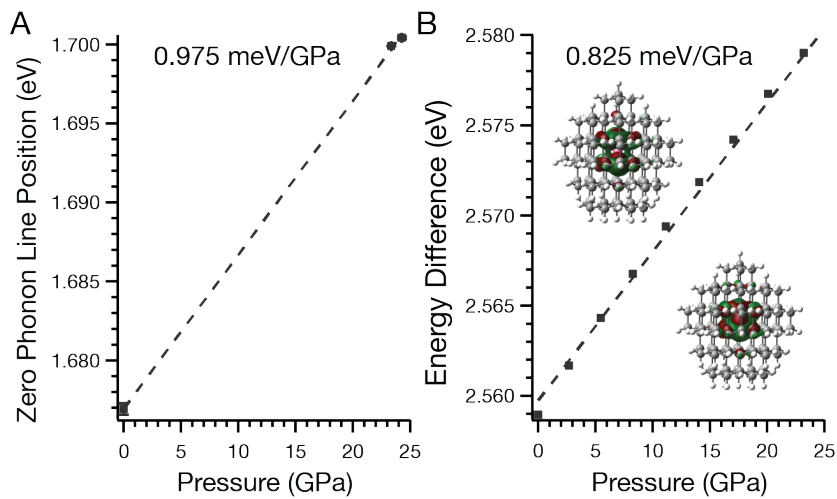

**Fig. 4. Evaluation of SiV⁻ center for optical pressure manometry.** (**A**) Experimental SiV⁻ ZPLs and (**B**) the B3LYP/6-31G(d) average energy differences of the molecular orbitals which exhibit largest contributions to the absorption peak responsible for the ZPL at different pressures. Error bars for both pressure and ZPL energy sit within the circular markers. The insets in (B) illustrate the contour plots (0.025 isodensity) of the LUMO (top) and the HOMO-2 (bottom) molecular orbitals (the largest contribution, see Tab. S1 and Fig. S6) of a SiV⁻ containing nanodiamond ($C_{119}SiH_{104}$), oriented perpendicular to the diamond <1,1,1> axis, as modeled with DFT. White, grey, and pink atoms are hydrogen, carbon, and silicon, respectively.